\documentclass[fleqn,usenatbib]{mnras}

\usepackage{newtxtext,newtxmath}

\usepackage[T1]{fontenc}

\DeclareRobustCommand{\VAN}[3]{#2}
\let\VANthebibliography\thebibliography
\def\thebibliography{\DeclareRobustCommand{\VAN}[3]{##3}\VANthebibliography}

\usepackage{graphicx}	
\usepackage{amsmath}	

\title[$\gamma$ rays from CRs escaping from GBH binaries]{Very-high-energy gamma rays from cosmic rays escaping from Galactic black hole binaries}

\author[Y. Ohira]{
Yutaka Ohira \thanks{E-mail: y.ohira@eps.s.u-tokyo.ac.jp}
\\
Department of Earth and Planetary Science, The University of Tokyo, 7-3-1 Hongo, Bunkyo-ku, Tokyo 113-0033, Japan\\
}

\date{Accepted 2025 July 15}
\pubyear{2025}

\begin{document}
\label{firstpage}
\pagerange{\pageref{firstpage}--\pageref{lastpage}}
\maketitle

\begin{abstract}
We solve the cosmic-ray diffusion around a Galactic black hole binary (microquasars) by considering the finite size of the escape region and the continuous cosmic-ray injection. 
We find that the energy spectrum of escaping cosmic rays in the gamma-ray emission region is described by a broken power law spectrum with one or two spectral breaks even though the total spectrum of escaping cosmic rays is a single power law spectrum. 
Using the solution for the diffusion equation, we construct a unified picture that explains spatially extended very-high-energy gamma rays from five microquasars observed by HAWC and LHAASO. 
The comparison of our unified model and observed data suggest that all five microquasars have the same energy spectrum of the escaping CRs, $dN/dE\propto E^{-2}$, the same diffusion coefficient, and the same emission region. 
The hard energy spectrum without the high-energy cutoff supports the idea that the origin of Galactic CRs beyond PeV energies is Galactic black hole binaries. 
\end{abstract}

\begin{keywords}
Cosmic rays -- stars: black holes -- diffusion
\end{keywords}



\section{Introduction}
The presence of cosmic rays (CRs) indicates that the present universe has a high-energy aspect.
Galactic CRs are thought to be accelerated by supernova remnants (SNRs) in our Galaxy. 
The gamma-ray telescopes AGILE and Fermi showed that Galactic CRs with energies below $1~{\rm TeV}$ are actually accelerated by Galactic SNRs \citep{2011ApJ...742L..30G,2013Sci...339..807A}. 
In the standard picture of Galactic CRs, protons and irons are accelerated to about $1~{\rm PeV}$ and $100~{\rm PeV}$, respectively.   
However, it is not yet understood theoretically and observationally to what energies SNRs can accelerate CRs. 
The TALE experiment shows that a non-negligible amount of CR protons are present even at $100~{\rm PeV}$ \citep{2021ApJ...909..178A}. 
Theoretically, a strong magnetic field amplification or some special SNRs are required to accelerate protons to $1~{\rm PeV}$ \citep{2004MNRAS.353..550B,2018MNRAS.478..926O,2022PhRvD.106l3025K,2024PhRvD.110d3046K}.
Moreover, the proton acceleration to $100~{\rm PeV}$ requires both the strong magnetic field and a special SNR \citep{2016A&A...595A..33T}. 

In addition to Galactic SNRs, black holes in our Galaxy have also been considered as CR sources \citep{2002A&A...390..751H,2017MNRAS.470.3332I,2020MNRAS.493.3212C}. 
It has been estimated that protons can be accelerated to $100~{\rm PeV}$ in the Galactic black hole binary system (so called, microquasar) \citep{2008A&A...485..623R,2020MNRAS.493.3212C,2020ApJ...904..188K,2021PASJ...73..530S}. 
In fact, very high energy (VHE) gamma rays with energies exceeding $100~{\rm TeV}$ from some microquasars was recently reported by the Large High Altitude Air Shower Observatory (LHAASO) and the High Altitude Water Cherenkov (HAWC) observatory  \citep{2018Natur.562...82A,2024Natur.634..557A,2024arXiv241008988L}. 
The spatial extent of the VHE gamma rays is several tens of pc, which is much larger than the size of a black hole binary system, but comparable to a hot bubble produced by the wind or jet from the black hole binary \citep{2021ApJ...910..149O}. 
Some microquasars have different gamma-ray spectra, some of which are steeper than predicted by the standard shock acceleration model \citep{1978MNRAS.182..147B,1978ApJ...221L..29B}.
Although individual interpretations for the VHE gamma rays from the microquasars SS 433 and V4641 Sgr are given by \citep{2018Natur.562...82A,2024Natur.634..557A,2024arXiv241008988L}, a unified explanation for the five microquasars that emit VHE gamma rays has not yet been attempted.

Gamma-ray spectra from middle-aged SNRs have similar characteristics: spatially extended gamma-ray distribution, different spectra, and steep spectra. 
These can be explained in a unified way by CRs escaping from SNRs \citep{2011MNRAS.410.1577O}. 
Gamma-ray spectra from escaping CRs are steeper than predicted by the standard shock acceleration model due to the energy-dependent diffusion \citep{1996A&A...309..917A,2009MNRAS.396.1629G}. 
In addition, the finite size of the escaping CR source has to be considered in order to understand the different gamma-ray spectra of some SNRs in a unified way \citep{2010MNRAS.409L..35L,2011MNRAS.410.1577O}. 

In this work, we try to understand the origin of the VHE gamma rays from microquasars in a unified way by considering CRs escaping from microquasars. 
For middle-aged SNRs, the CR acceleration to the TeV-PeV scale has already completed, so that the injection of escaping CRs can be treated as an impulsive injection. 
On the other hand, as long as the black hole continues to accrete mass, the microquasar would continue to accelerate CRs. 
Therefore, escaping CRs would be continuously injected into the surroundings. 
First, we derive the distribution of escaping CRs that are continuously injected in the finite size region. 
Then, we show that VHE gamma-ray spectra from five microquasars can be explained in a unified way by CRs escaping from the microquasars.

\section{Energy spectrum of escaping CRs in a gamma-ray emitting region}
We consider a spherically symmetric system for simplicity. 
A microquasar launches a pair of jets or wind, creating a low-density hot bubble with the size of 
\begin{equation}
R_1= 78~{\rm pc}~ \left(\frac{L}{10^{39}~{\rm erg~s^{-1}}}\right)^{1/5}\left(\frac{n}{10~{\rm cm^{-3}}}\right)^{-1/5}\left(\frac{t_{\rm age}}{10^6~{\rm yr}}\right)^{3/5}~~,
\end{equation}
where $L, n$, and $t_{\rm age}$ are the outflow luminosity, density of the surroundings, and age of the object, respectively \citep{1975ApJ...200L.107C,1977ApJ...218..377W}. 
If the total outflow energy from the microquasar is smaller than the energy of the 
supernova explosion that made the black hole, the bubble around the microquasar 
would be generated by the supernova remnant \citep{2017MNRAS.468.1226G}. 
CRs accelerated in the microquasar system are initially trapped in the bubble, but eventually escape from there. 
Therefore, gamma rays are expected to be produced by pp interaction outside the low-density hot bubble. 

We set $R_1$ and $R_2$ as the inner and outer radii of the gamma-ray emitting region, that is,  the VHE gamma rays are produced in $R_1\leq r\leq R_2$. 
$R_2$ corresponds to the radius of the forward shock driven by the hot bubble or the outer radius of a dense region (atomic or molecular cloud). 
As a first step, $R_1$ and $R_2$ are assumed to be constant in this work. 
This assumption is justified in the following cases. 

If there is a dense region and the mass swept up by the forward shock is smaller than the total mass of the dense region, $R_2$ should be the outer radius of the dense region. 
Then, $R_2$ does not depend on time. 
It takes a finite time $t_{\rm esc}$ for CRs to propagate from the acceleration region in the bubble to the boundary of the bubble, $R_1$. 
Thus, CRs start escaping from the bubble at $t_{\rm esc}$. 
For $t_{\rm age} - t_{\rm esc} \ll t_{\rm age}$, the bubble size, $R_1$, does not significantly change during $t_{\rm age} - t_{\rm esc}$.
In addition, it takes another finite time, $t_{\rm amp}$, to amplify magnetic field fluctuations in the emission region by the CR streaming instability \citep{1967ApJ...147..689L,1969ApJ...156..445K}. 
Before this time ($t_{\rm age}<t_{\rm amp}$), CRs quickly escape from the emission region because the diffusion coefficient is large. 
However, once the magnetic field fluctuations are sufficiently amplified by the CR streaming instability, the diffusion coefficient becomes small, so that a lot of CRs are confined in the emission region \citep{2010ApJ...712L.153F,2013ApJ...768...73M}. 
In this case, CRs injected for $t_{\rm age} - t_{\rm amp}$ mainly contribute gamma rays from the emission region. 
For $t_{\rm age} - t_{\rm amp}\ll t_{\rm age}$, the bubble size, $R_1$, does not significantly change again during $t_{\rm age} - t_{\rm amp}$.
The magnetic field strength in the bubble is expected to depend on time \citep{1973ApJ...186..249P}. 
Recent fluid simulations show that magnetic fields are very turbulent in the shocked wind (or jet) region \citep{2014MNRAS.438..278P,2023A&A...679A..49M}. 
However, the time evolution of the magnetic turbulence in the bubble and emission regions is still open question. 
Understanding $t_{\rm esc}$ in the bubble and $t_{\rm amp}$ in the emission region is beyond the scope of this work, but it should be addressed in the future.

After escaping from the hot bubble region, the escaping CRs diffusively propagate to the surroundings. 
We derive the distribution of the escaping CRs, $f(t,r,p)$, at distance $r$ from the microquasar center and at time $t$, where $p$ is the CR momentum. 
Note that $t$ is the time since the CRs started to escape or the time since the magnetic field turbulence started to be amplified in the gamma-ray emitting region. The Green function of the diffusion equation in a spherically symmetric system with a uniform diffusion coefficient is given by
\begin{equation}
G(t-t',|{\bf r}-{\bf r'}|,p)=\frac{e^{-\left(\frac{|{\bf r}-{\bf r'}|^2}{R_{\rm d}(t-t',p)}\right)^2}}{\pi^{3/2}R_{\rm d}(t-t',p)^3} ~~,
\label{eq:fpoint}
\end{equation}
where 
\begin{equation}
R_{\rm d}(t-t',p) = \sqrt{4D_{\rm xx}(p)(t-t')}
\label{eq:rd}
\end{equation}
is the diffusion length and $D_{\rm xx}(p)$ is the diffusion coefficient, which is assumed to 
depend only on the CR momentum in this work. 
We assume that CRs are continuously escaping from the bubble radius, $R_1$. 
Then, the source term in the diffusion equation is given by 
\begin{equation}
q_{\rm s}(t,r,p)=\frac{Q_{\rm esc}(p)}{4\pi r^2} \delta(r-R_1) ~~,
\end{equation}
where $Q_{\rm esc}(p)$ is the escaping CR spectrum injected per time. 
Then, the solution to the diffusion equation is given by 
\begin{equation}
f(t,r,p)=\frac{Q_{\rm esc}(p)}{4\pi^{3/2}R_1r}  \int_0^t {\rm d}t'  \frac{e^{-\left(\frac{r-R_1}{R_{\rm d}(t-t',p)}\right)^2}-e^{-\left(\frac{r+R_1}{R_{\rm d}(t-t',p)}\right)^2}}  {R_{\rm d}(t-t',p)}~~.
\label{eq:f}
\end{equation}
Since the above integrand is approximately $1/R_{\rm d}$ for $r\sim R_1 > R_{\rm d}$, but $4R_1r\exp\{-(r/R_{\rm d})^2\}/R_{\rm d}^3$ for the others \citep{2011MNRAS.410.1577O}, 
the approximate solution to the diffusion equation is 
\begin{equation}
f(t,r,p)=\frac{Q_{\rm esc}(p)t}{2\pi^{3/2}R_1R_{\rm d}(t,p)r}  
\label{eq:af1}
\end{equation}
for $r\sim R_1 > R_{\rm d}$, and 
\begin{equation}
f(t,r,p)=\frac{Q_{\rm esc}(p)t}{\pi R_{\rm d}(t,p)^2r} {\rm erfc}\left(\frac{r}{R_{\rm d}(t,p)}\right) \label{eq:af2}
\end{equation}
for the others, where ${\rm erfc}(x) = (2/\sqrt{\pi})\int_x^{\infty}e^{-y^2}{\rm d}y$ is the complementary error function. 

The spectrum of escaping CRs in the emission region $R_1\leq r\leq R_2$ is given by
\begin{eqnarray}
F(t,p)&=& \int_{R_1}^{R_2} f(t,r,p)4\pi r^2{\rm d}r \nonumber \\
&=&Q_{\rm esc}(p)t M(t,p)
\label{eq:fp}
\end{eqnarray}
where $M(t,p)$ is given by 
%
\begin{eqnarray}
&M(t,p)= \frac{1}{2t}\int_0^t {\rm d}t' \left[\frac{R_{\rm d}(t-t',p)}{\sqrt{\pi}R_1}\left\{ 1 -e^{- \left( \frac{R_2-R_1}{R_{\rm d}(t-t',p)} \right)^2} \right. \right. ~~~~~~~~~~~~&  \nonumber \\
& ~~~~~~~~~~~~~~~~~~~~~~~~~~~~~~~~~~~\left.  +e^{- \left( \frac{R_2+R_1}{R_{\rm d}(t-t',p)} \right)^2} -e^{- \left( \frac{2R_1}{R_{\rm d}(t-t',p)} \right)^2} \right\} & \nonumber \\ 
&~~~~~~\left.+{\rm erf}\left(\frac{R_2-R_1}{R_{\rm d}(t-t',p)}\right) +{\rm erf}\left(\frac{R_2+R_1}{R_{\rm d}(t-t',p)}\right)- {\rm erf}\left(\frac{2R_1}{R_{\rm d}(t-t',p)}\right) \right]&.
\label{eq:mp}
\end{eqnarray}
%
${\rm erf}(x)=(2/\sqrt{\pi})\int_0^{x}e^{-y^2}{\rm d}y$ is the error function.
If all escaping CRs are in the emission region, $F(p,t) = Q_{\rm esc}(p) t$. 
Therefore, the function, $M(p,t)$, describes the spectral modification due to the diffusive propagation of escaping CRs, which depends only on $R_1/R_{\rm d}(t,p)$ and $R_2/R_1$.

For $R_2/R_1\sim 1$, since the emission region is close to the escape radius, the finite size of the escape region affects the spectrum of escaping CRs in the emission region, $F(t,p)$.  
For $R_{\rm d}(t,p) < R_2-R_1\lesssim R_1$, since about half of escaping CRs are in the emission region, $M(t,p)$ is approximately given by
\begin{equation}
M(t,p)\approx 0.5~.
\label{eq:am1}
\end{equation}
For $R_2-R_1<R_{\rm d}(t,p)<R_1$, CRs can diffuse beyond the emission region but cannot diffuse far from the escape radius, $R_1$.
Since propagation from $R_1$ can be described by one-dimensional diffusion and equation~(\ref{eq:af1}) can be applied, $M(t,p)$ is approximately given by
\begin{equation}
M(t,p)\approx \frac{R_2^2 -R_1^2}{2R_1\sqrt{\pi D_{\rm xx}(p)t} }~~.
\label{eq:am2}
\end{equation}
For $R_1<R_{\rm d}(t,p)$, since propagation from $R_1$ can be described by three-dimensional diffusion and equation~(\ref{eq:af2}) can be applied, $M(t,p)$ is approximately given by
\begin{equation}
M(t,p)\approx \frac{R_2^2 -R_1^2}{2D_{\rm xx}(p)t}~~.
\label{eq:am3}
\end{equation}
Therefore, the spectrum of escaping CRs in the emission region, $F(t,p)$, has two spectral breaks at $p_{2-1}$ and $p_1$ for $R_1\sim R_2$, 
where $p_{2-1}$ and $p_1$ are given by the conditions $R_{\rm d}(t,p) = R_2-R_1$ and $R_{\rm d}(t,p)=R_1$, respectively. 
For $p>p_1$, the steady state is realized, so that the CR spectrum in the emission region does not depend on $t$.

For $R_2/R_1\gg 1$, the source region can be regarded as the point source because the escape region is much smaller than the emission region. 
In this case, from equation~(\ref{eq:af2}), $M(t,p)$ is approximately given by
\begin{eqnarray}
M(t,p) =  \left\{ \begin{array}{ll}
1& ({\rm for}~R_{\rm d}(t,p)<R_2) \\
\frac{R_2^2 -R_1^2}{2D_{\rm xx}(p)t} & ({\rm for}~R_2<R_{\rm d}(t,p)) 
\end{array} \right. ~~.
\label{eq:am2}
\end{eqnarray}
Thus, the spectrum of escaping CRs in the emission region, $F(t,p)$, has one spectral break at $p_2$, where $p_2$ is given by the condition $R_{\rm d}(t,p) = R_2$.

If the momentum dependences of the source spectrum and diffusion coefficient are given by $Q_{\rm esc}(p)\propto p^{-s_{\rm esc}}$ and $D_{\rm xx}(p)\propto p^{\delta}$, for $R_2/R_1\sim 1$, the spectral index of escaping CRs in the emission region is approximately given by 
\begin{eqnarray}
\frac{d\log F(t,p)}{d\log p} \propto  \left\{ \begin{array}{lll}
-s_{\rm esc}& (p<p_{2-1}) \\
-(s_{\rm esc}+\delta/2) & (p_{2-1}<p<p_1) \\
-(s_{\rm esc}+\delta)  & (p_1<p) \\
\end{array} \right. , 
\label{eq:af3}
\end{eqnarray}
and for $R_2/R_1\gg 1$, 
\begin{eqnarray}
\frac{d\log F(t,p)}{d\log p} \propto  \left\{ \begin{array}{ll}
-s_{\rm esc}& (p<p_2) \\
-(s_{\rm esc}+\delta)  & (p_2<p) 
\end{array} \right. ~~.
\label{eq:af4}
\end{eqnarray}
The spectral index in the hardest momentum region represents the true spectral index of the total escaping cosmic rays, which is not always the same as the spectral index in the acceleration region \citep{2010A&A...513A..17O,2011ApJ...729L..13O}. 

\begin{figure}
\centering
\includegraphics[width=\columnwidth]{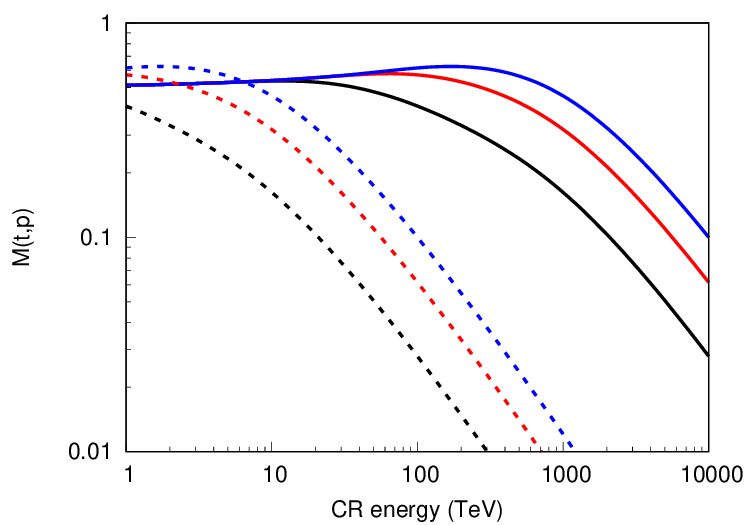}
\caption{The black, red, and blue lines show $M(t,p)$ for $R_1=30~{\rm pc}, R_2=40, 50, 60~{\rm pc}$, respectively. The solid and dashed lines show $M(t,p)$ for $t=10^5, 10^7~{\rm yr}$, respectively. The diffusion coefficient is the Bohm diffusion ($\delta=1$) in $10~ {\rm \mu G}$.}
\label{fig1}
\end{figure}

Fig.~\ref{fig1} shows $M(p,t)$ for some cases, where the diffusion coefficient, $D_{\rm xx}$, is assumed to be the Bohm diffusion ($\delta=1$) in $10~ {\rm \mu G}$, $t=10^5$ and $10^7~{\rm yr}, R_1=30~{\rm pc}$, and $R_2 = 40, 50$, and $60 ~{\rm pc}$. 
For the Bohm diffusion coefficient, the characteristic break energy is estimated by
\begin{equation}
cp_{\rm b} =0.2~{\rm PeV}\left(\frac{B}{10~{\rm \mu G}}\right)\left(\frac{R}{30~{\rm pc}}\right)^2\left(\frac{t}{10^5~{\rm yr}}\right)^{-1}~,
\label{eq:ebreak}
\end{equation}
where $R$ is $R_2-R_1, R_1$, or $R_2$. 
As can be seen in Fig.~\ref{fig1}, there are two characteristic breaks for $R_2/R_1 \sim 1$ ($R_2=40$ and $50 {\rm pc}$), and one characteristic break  for $R_2/R_1>1$ ($R_2=60~{\rm pc}$).
The break energies and spectral indices are consistent with equations~(\ref{eq:af3})-(\ref{eq:ebreak}). 

Note that we have not taken account of the cutoff in the low energy side. 
If the inner radius of the emission region is larger than the bubble radius, low energy CRs with a small diffusion coefficient cannot propagate to the emission region for a finite time. 
As a result, the gamma-ray spectrum has a cutoff structure in the low energy side \citep{1996A&A...309..917A,2009MNRAS.396.1629G,2011MNRAS.410.1577O}.

\section{Gamma-ray spectra from cosmic rays escaping from microquasars}
\begin{figure}
\includegraphics[width=\columnwidth]{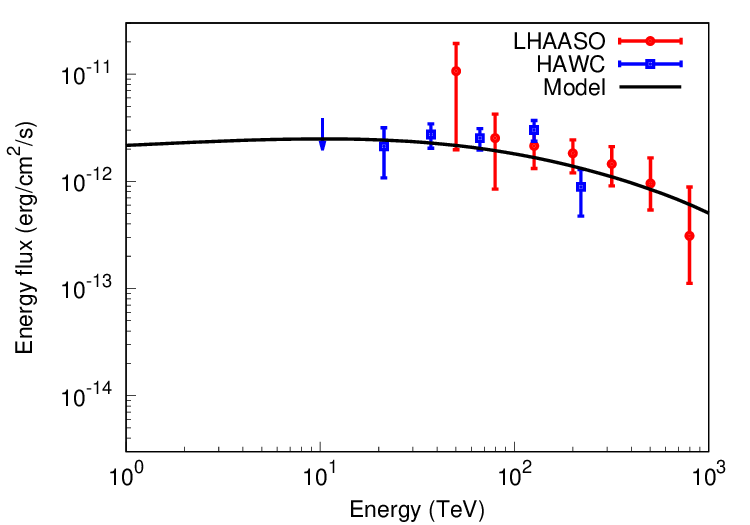}
\caption{Comparison of the model (solid line) with LHAASO (red) and HWAC observations for V4641 Sgr.}
\label{fig2} 
\end{figure}
\begin{figure}
\includegraphics[width=\columnwidth]{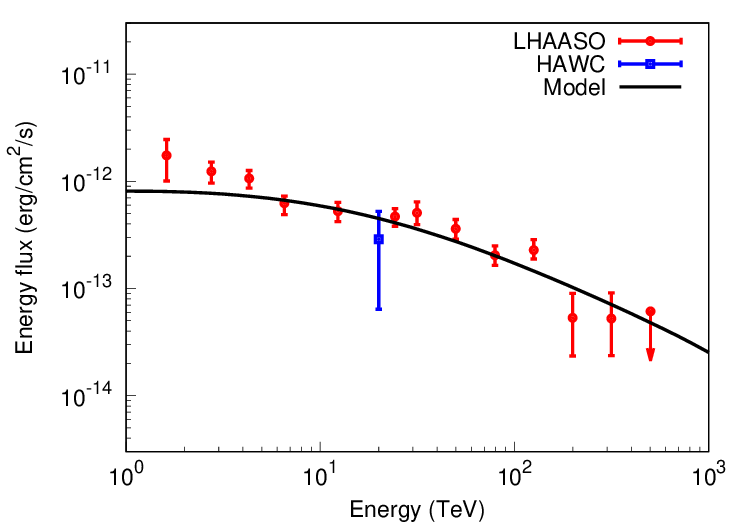}
\caption{Same as Fig.~\ref{fig2}, but for SS 433.}
\label{fig3} 
\end{figure}
\begin{figure}
\includegraphics[width=\columnwidth]{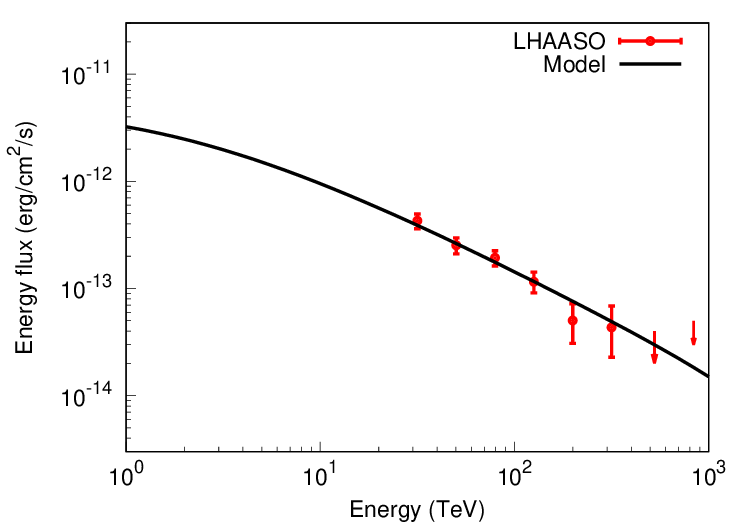}
\caption{Same as Fig.~\ref{fig2}, but for GRS 1915+105.}
\label{fig4} 
\end{figure}
\begin{figure}
\includegraphics[width=\columnwidth]{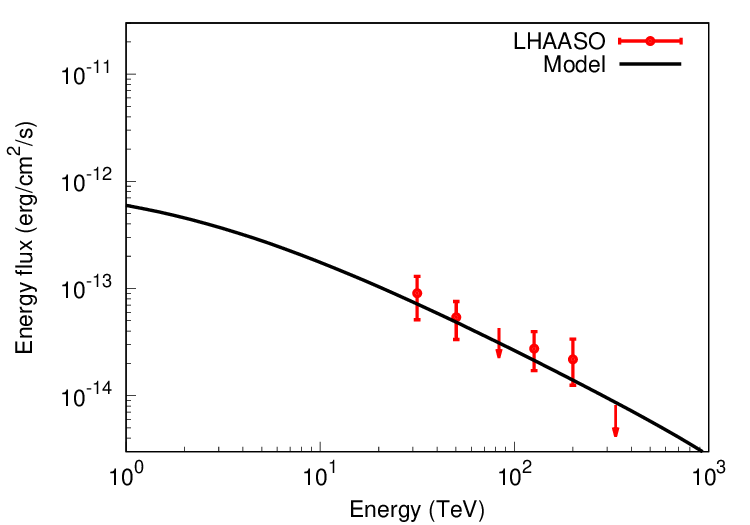}
\caption{Same as Fig.~\ref{fig2}, but for MAXI J1820+070.}
\label{fig5} 
\end{figure}
\begin{figure}
\includegraphics[width=\columnwidth]{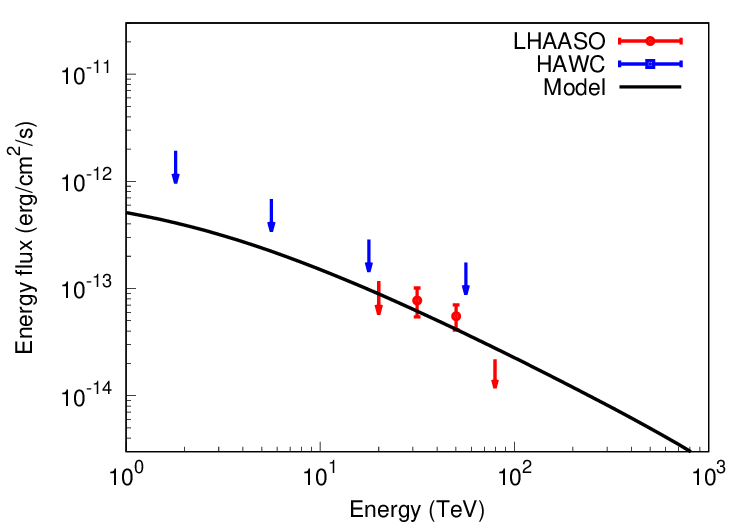}
\caption{Same as Fig.~\ref{fig2}, but for Cygnus X-1.}
\label{fig6} 
\end{figure}

Gamma-ray spectra from escaping CRs are calculated by using equations~(\ref{eq:rd}), (\ref{eq:fp}) and (\ref{eq:mp}) and  the Naima python package \citep{2015ICRC...34..922Z}, in which gamma rays are calculated by the cross section of pp interactions given in \citet{2014PhRvD..90l3014K}. 
By comparing our model with the observational data provided by LAHHSO and HAWC, we try to find a unified picture that the extended VHE gamma rays from five microquasars are produced by CRs escaping from the microquasars, rather than finding the best fitting parameters for each of the microquasars. 
The gamma-ray flux depends on the CR luminosity, distance, and total mass in the emission region. However, it is very difficult to estimate these values accurately. 
In this work, we adjust the normalization of the gamma-ray flux to fit the data.

Figs.~\ref{fig2}-\ref{fig6} show the VHE gamma-ray spectra for V4641 Sgr, SS 433, GRS 1915+105, MAXI J1820+070, and Cygnus X-1, respectively. 
Our model is in broad agreement with observational data. 
We find a unified picture that all of these objects have the same parameter values, except for their time since CRs started to escape, $t$. 
In the unified model, the total energy spectrum of escaping CRs is $E^{-2}$, the diffusion coefficient is the Bohm diffusion coefficient ($\delta=1$) in $10~{\rm \mu G}$, $R_1=30~{\rm pc}$ and $R_2=60~{\rm pc}$. 
The time of these objects are assumed to be $t=10^5~{\rm yr}$ for V4641 Sgr, $t=10^6~{\rm yr}$ for SS433, and $t=10^7~{\rm yr}$ for GRS 1915+105, MAXI J1820+070, Cygnus X-1. 
Therefore, the time of the objects, $t$, is responsible for the spectral differences between them in the unified model. 
This interpretation of the VHE gamma-rays from microquasars is not unique because the spectral structure depends only on $R_1/R_{\rm d}(t,p)$ and $R_2/R_{\rm d}(t,p)$.
Thus, the radius or the diffusion coefficient may also be responsible for the spectral difference. 

Characteristic break energies in the gamma-ray spectrum can be estimated from equation~(\ref{eq:ebreak}) as $E_{\rm \gamma,b}\approx 0.1cp_{\rm b}$. 
Since the characteristic radii are $R_1=R_2-R_1=30~{\rm pc}$ and $R_2=60~{\rm pc}$ in our unified model, the break energy is mainly characterized by the outer radius, that is, $p_{\rm b}= p_2$, where $p_2$ is given by $R_{\rm d}(t,p)=R_2$. 
Then, the characteristic break energies are estimated as $E_{\rm \gamma,b} \approx 80~{\rm TeV}$ for V4641 Sgr, $8~{\rm TeV}$ for SS433, and $0.8~{\rm TeV}$ for GRS 1915+105, MAXI J1820+070, Cygnus X-1, respectively. 
These values are consistent with the observed gamma-ray spectra. 

Interestingly, the comparison of our unified model and observed data suggests that for all microquasars, the total energy spectra of escaping CRs are $dN/dE\propto E^{-2}$ with no cutoff at the high energy end. 
Since this spectrum is harder than that in SNRs \citep{2010A&A...513A..17O}, microquasars have the potential to supply Galactic cosmic rays above the PeV scale. 
These interesting suggestions are not derived theoretically in this work. Further investigation is needed to understand how microquasars produce the energy spectrum of $dN/dE\propto E^{-2}$ and accelerate particles beyond the PeV scale.

\section{Discussion} 
To explain the hard spectrum observed in V4641 Sgr and SS 433, 
$R_1/R_{\rm d}=R_1(4D_{\rm xx}t)^{-1/2}$ should be about $0.5$ and $0.15$ at energy of $cp=1~{\rm PeV}$. 
Thus, the diffusion coefficient at energy of $cp=1~{\rm PeV}$ should be
\begin{equation}
D_{\rm xx} =2.7\times 10^{27} {\rm cm^{2} s^{-1}} \left(\frac{R_1}{30~{\rm pc}}\right)^2 \left(\frac{t}{10^5~{\rm yr}}\right)^{-1}\left(\frac{R_1/R_{\rm d}}{0.5}\right)^{-2}.
\label{eq:dxx}
\end{equation}
This value is the same as the Bohm diffusion coefficient in $10~{\rm \mu G}$ and smaller than the Galactic mean value expected from the CR boron-to-carbon ratio \citep{2024Natur.634..557A,2024arXiv241008988L}. 
The smaller diffusion coefficient around the CR sources is also expected from gamma-ray observations of SNRs \citep{2008MNRAS.387L..59T,2009ApJ...707L.179F}. 
Escaping CRs are expected to amplify the magnetic field turbulence through the CR streaming instability, 
so that the diffusion coefficient could be small around CR sources \citep{2010ApJ...712L.153F,2013ApJ...768...73M}. 
However, to make the diffusion coefficient the Bohm limit, the amplitude of magnetic field turbulence has to be amplified to the background magnetic field strength, which is larger than expected around SNRs. 
Compared to escaping CRs around SNRs, escaping CRs can keep the diffusive flux sufficiently large around microquars for a long time because of the continuous injection for a longer time. 
Then, the magnetic field turbulence could be amplified strongly compared with SNRs. 
In addition, the gyroradius of PeV protons is $0.1~{\rm pc}(B/10~{\rm \mu G})^{-1}$ which is larger than that of GeV-TeV protons around SNRs. 
Therefore, other mechanisms, such as Rayleigh-Taylor or thermal instabilities, may be responsible for generating the magnetic field turbulence around microquasars. 
The confinement of PeV CRs around microquasars should be studied in detail elsewhere. 

In addition to CR protons, microquasars would accelerate CR electrons. 
Then, the CR electrons could escape from the microquasar system, producing extended gamma rays through inverse Compton scattering \citep{2012MNRAS.427...91O}. 
The propagation of CR electrons is different from that of protons because CR electrons suffer radiative losses. 
If we observe gamma rays from both CR electrons and protons and distinguish spatially, 
the diffusion coefficient around CR sources can be estimated from gamma-ray observations. 
We look forward to further observations by LHAASO and HAWC, and the future new gamma-ray and VHE gamma-ray observations, Cherenkov Telescope Array (CTA), Andes Large-area PArticle detector for Cosmic-ray physics and Astronomy (ALPACA), the Southern Wide-field Gamma-ray Observatory(SWGO).

In this work, we have not investigated where and how cosmic rays are accelerated in the microquasar system. 
Many acceleration mechanisms are expected to work: shock acceleration in jet and wind \citep{1978MNRAS.182..147B,1978ApJ...221L..29B,2024ApJ...969L...1M}, turbulent acceleration in the hot bobble \citep{2009MNRAS.400..248O,2013ApJ...767L..16O}, shear acceleration at the jet edge \citep{1998A&A...335..134O,2004ApJ...617..155R,2018PhRvD..97b3026K}, and so on. 
However, we do not understand which process mainly accelerates CRs in the microquasar system. 
The detailed physics of escape and acceleration of CRs should be investigated simultaneously. 
Since the microquasar system can be regarded as a smaller version of the active galactic nuclei system, 
we could understand the acceleration and escape processes of extragalactic CRs by observing the microquasar system. 
In terms of the steady-state injection of CRs, CRs accelerated in pulsar wind nebulae and stellar wind may also share common physics. 
Therefore, understanding CRs from microquasars is useful for understanding the high-energy aspect of the present universe.

\section{Summary} 
HAWC and LHASSO reported spatially extended VHE gamma rays from five Galactic black hole binaries (microquasars). 
The large emission region suggests that the VHE gamma rays originate from CRs escaping from the microquasars. 
To understand the distribution of escaping CRs around a microquasar, we have solved the diffusion equation taking into account the finite size of the CR source and the continuous CR injection. 
The energy spectrum in the emission region is described by a broken power law spectrum with one or two spectral breaks even though the total spectrum of escaping CRs is a single power law spectrum. 
By comparing our solution with the observed data, we have found the unified picture that all five microquasars have the same energy spectrum of the escaping CRs, $dN/dE\propto E^{-2}$, the same diffusion coefficient, and the same emission region. 
The hard energy spectrum of escaping CRs without the high energy cutoff supports the idea that the origin of Galactic CRs above the PeV energy is Galactic microquasars.

\section*{Acknowledgements}
This work is supported by JSPS KAKENHI grant Nos. JP21H04487, JP24H01805 and JP25K00999.

\section*{Data Availability}
The data underlying this work will be shared on reasonable request to the corresponding author.



\bibliographystyle{mnras}
\bibliography{ref.bib} 

\begin{thebibliography}{}
\makeatletter
\relax
\def\mn@urlcharsother{\let\do\@makeother \do\$\do\&\do\#\do\^\do\_\do\%\do\~}
\def\mn@doi{\begingroup\mn@urlcharsother \@ifnextchar [ {\mn@doi@}
  {\mn@doi@[]}}
\def\mn@doi@[#1]#2{\def\@tempa{#1}\ifx\@tempa\@empty \href
  {http://dx.doi.org/#2} {doi:#2}\else \href {http://dx.doi.org/#2} {#1}\fi
  \endgroup}
\def\mn@eprint#1#2{\mn@eprint@#1:#2::\@nil}
\def\mn@eprint@arXiv#1{\href {http://arxiv.org/abs/#1} {{\tt arXiv:#1}}}
\def\mn@eprint@dblp#1{\href {http://dblp.uni-trier.de/rec/bibtex/#1.xml}
  {dblp:#1}}
\def\mn@eprint@#1:#2:#3:#4\@nil{\def\@tempa {#1}\def\@tempb {#2}\def\@tempc
  {#3}\ifx \@tempc \@empty \let \@tempc \@tempb \let \@tempb \@tempa \fi \ifx
  \@tempb \@empty \def\@tempb {arXiv}\fi \@ifundefined
  {mn@eprint@\@tempb}{\@tempb:\@tempc}{\expandafter \expandafter \csname
  mn@eprint@\@tempb\endcsname \expandafter{\@tempc}}}

\bibitem[\protect\citeauthoryear{{Abbasi} et~al.,}{{Abbasi}
  et~al.}{2021}]{2021ApJ...909..178A}
{Abbasi} R.~U.,  et~al., 2021, \mn@doi [\apj] {10.3847/1538-4357/abdd30}, \href
  {https://ui.adsabs.harvard.edu/abs/2021ApJ...909..178A} {909, 178}

\bibitem[\protect\citeauthoryear{{Abeysekara} et~al.,}{{Abeysekara}
  et~al.}{2018}]{2018Natur.562...82A}
{Abeysekara} A.~U.,  et~al., 2018, \mn@doi [\nat] {10.1038/s41586-018-0565-5},
  \href {https://ui.adsabs.harvard.edu/abs/2018Natur.562...82A} {562, 82}

\bibitem[\protect\citeauthoryear{{Ackermann} et~al.,}{{Ackermann}
  et~al.}{2013}]{2013Sci...339..807A}
{Ackermann} M.,  et~al., 2013, \mn@doi [Science] {10.1126/science.1231160},
  \href {https://ui.adsabs.harvard.edu/abs/2013Sci...339..807A} {339, 807}

\bibitem[\protect\citeauthoryear{{Aharonian} \& {Atoyan}}{{Aharonian} \&
  {Atoyan}}{1996}]{1996A&A...309..917A}
{Aharonian} F.~A.,  {Atoyan} A.~M.,  1996, \aap, \href
  {https://ui.adsabs.harvard.edu/abs/1996A&A...309..917A} {309, 917}

\bibitem[\protect\citeauthoryear{{Alfaro} et~al.,}{{Alfaro}
  et~al.}{2024}]{2024Natur.634..557A}
{Alfaro} R.,  et~al., 2024, \mn@doi [\nat] {10.1038/s41586-024-07995-9}, \href
  {https://ui.adsabs.harvard.edu/abs/2024Natur.634..557A} {634, 557}

\bibitem[\protect\citeauthoryear{{Bell}}{{Bell}}{1978}]{1978MNRAS.182..147B}
{Bell} A.~R.,  1978, \mn@doi [\mnras] {10.1093/mnras/182.2.147}, \href
  {http://adsabs.harvard.edu/abs/1978MNRAS.182..147B} {182, 147}

\bibitem[\protect\citeauthoryear{{Bell}}{{Bell}}{2004}]{2004MNRAS.353..550B}
{Bell} A.~R.,  2004, \mn@doi [\mnras] {10.1111/j.1365-2966.2004.08097.x}, \href
  {http://adsabs.harvard.edu/abs/2004MNRAS.353..550B} {353, 550}

\bibitem[\protect\citeauthoryear{{Blandford} \& {Ostriker}}{{Blandford} \&
  {Ostriker}}{1978}]{1978ApJ...221L..29B}
{Blandford} R.~D.,  {Ostriker} J.~P.,  1978, \mn@doi [\apjl] {10.1086/182658},
  \href {http://adsabs.harvard.edu/abs/1978ApJ...221L..29B} {221, L29}

\bibitem[\protect\citeauthoryear{{Castor}, {McCray}  \& {Weaver}}{{Castor}
  et~al.}{1975}]{1975ApJ...200L.107C}
{Castor} J.,  {McCray} R.,   {Weaver} R.,  1975, \mn@doi [\apjl]
  {10.1086/181908}, \href
  {https://ui.adsabs.harvard.edu/abs/1975ApJ...200L.107C} {200, L107}

\bibitem[\protect\citeauthoryear{{Cooper}, {Gaggero}, {Markoff}  \&
  {Zhang}}{{Cooper} et~al.}{2020}]{2020MNRAS.493.3212C}
{Cooper} A.~J.,  {Gaggero} D.,  {Markoff} S.,   {Zhang} S.,  2020, \mn@doi
  [\mnras] {10.1093/mnras/staa373}, \href
  {https://ui.adsabs.harvard.edu/abs/2020MNRAS.493.3212C} {493, 3212}

\bibitem[\protect\citeauthoryear{{Fujita}, {Ohira}, {Tanaka}  \&
  {Takahara}}{{Fujita} et~al.}{2009}]{2009ApJ...707L.179F}
{Fujita} Y.,  {Ohira} Y.,  {Tanaka} S.~J.,   {Takahara} F.,  2009, \mn@doi
  [\apjl] {10.1088/0004-637X/707/2/L179}, \href
  {https://ui.adsabs.harvard.edu/abs/2009ApJ...707L.179F} {707, L179}

\bibitem[\protect\citeauthoryear{{Fujita}, {Ohira}  \& {Takahara}}{{Fujita}
  et~al.}{2010}]{2010ApJ...712L.153F}
{Fujita} Y.,  {Ohira} Y.,   {Takahara} F.,  2010, \mn@doi [\apjl]
  {10.1088/2041-8205/712/2/L153}, \href
  {https://ui.adsabs.harvard.edu/abs/2010ApJ...712L.153F} {712, L153}

\bibitem[\protect\citeauthoryear{{Gabici}, {Aharonian}  \& {Casanova}}{{Gabici}
  et~al.}{2009}]{2009MNRAS.396.1629G}
{Gabici} S.,  {Aharonian} F.~A.,   {Casanova} S.,  2009, \mn@doi [\mnras]
  {10.1111/j.1365-2966.2009.14832.x}, \href
  {https://ui.adsabs.harvard.edu/abs/2009MNRAS.396.1629G} {396, 1629}

\bibitem[\protect\citeauthoryear{{Giuliani} et~al.,}{{Giuliani}
  et~al.}{2011}]{2011ApJ...742L..30G}
{Giuliani} A.,  et~al., 2011, \mn@doi [\apjl] {10.1088/2041-8205/742/2/L30},
  \href {https://ui.adsabs.harvard.edu/abs/2011ApJ...742L..30G} {742, L30}

\bibitem[\protect\citeauthoryear{{Grichener} \& {Soker}}{{Grichener} \&
  {Soker}}{2017}]{2017MNRAS.468.1226G}
{Grichener} A.,  {Soker} N.,  2017, \mn@doi [\mnras] {10.1093/mnras/stx534},
  \href {https://ui.adsabs.harvard.edu/abs/2017MNRAS.468.1226G} {468, 1226}

\bibitem[\protect\citeauthoryear{{Heinz} \& {Sunyaev}}{{Heinz} \&
  {Sunyaev}}{2002}]{2002A&A...390..751H}
{Heinz} S.,  {Sunyaev} R.,  2002, \mn@doi [\aap] {10.1051/0004-6361:20020615},
  \href {https://ui.adsabs.harvard.edu/abs/2002A&A...390..751H} {390, 751}

\bibitem[\protect\citeauthoryear{{Ioka}, {Matsumoto}, {Teraki}, {Kashiyama}  \&
  {Murase}}{{Ioka} et~al.}{2017}]{2017MNRAS.470.3332I}
{Ioka} K.,  {Matsumoto} T.,  {Teraki} Y.,  {Kashiyama} K.,   {Murase} K.,
  2017, \mn@doi [\mnras] {10.1093/mnras/stx1337}, \href
  {https://ui.adsabs.harvard.edu/abs/2017MNRAS.470.3332I} {470, 3332}

\bibitem[\protect\citeauthoryear{{Kafexhiu}, {Aharonian}, {Taylor}  \&
  {Vila}}{{Kafexhiu} et~al.}{2014}]{2014PhRvD..90l3014K}
{Kafexhiu} E.,  {Aharonian} F.,  {Taylor} A.~M.,   {Vila} G.~S.,  2014, \mn@doi
  [\prd] {10.1103/PhysRevD.90.123014}, \href
  {https://ui.adsabs.harvard.edu/abs/2014PhRvD..90l3014K} {90, 123014}

\bibitem[\protect\citeauthoryear{{Kamijima} \& {Ohira}}{{Kamijima} \&
  {Ohira}}{2022}]{2022PhRvD.106l3025K}
{Kamijima} S.~F.,  {Ohira} Y.,  2022, \mn@doi [\prd]
  {10.1103/PhysRevD.106.123025}, \href
  {https://ui.adsabs.harvard.edu/abs/2022PhRvD.106l3025K} {106, 123025}

\bibitem[\protect\citeauthoryear{{Kamijima} \& {Ohira}}{{Kamijima} \&
  {Ohira}}{2024}]{2024PhRvD.110d3046K}
{Kamijima} S.~F.,  {Ohira} Y.,  2024, \mn@doi [\prd]
  {10.1103/PhysRevD.110.043046}, \href
  {https://ui.adsabs.harvard.edu/abs/2024PhRvD.110d3046K} {110, 043046}

\bibitem[\protect\citeauthoryear{{Kimura}, {Murase}  \& {Zhang}}{{Kimura}
  et~al.}{2018}]{2018PhRvD..97b3026K}
{Kimura} S.~S.,  {Murase} K.,   {Zhang} B.~T.,  2018, \mn@doi [\prd]
  {10.1103/PhysRevD.97.023026}, \href
  {https://ui.adsabs.harvard.edu/abs/2018PhRvD..97b3026K} {97, 023026}

\bibitem[\protect\citeauthoryear{{Kimura}, {Murase}  \&
  {M{\'e}sz{\'a}ros}}{{Kimura} et~al.}{2020}]{2020ApJ...904..188K}
{Kimura} S.~S.,  {Murase} K.,   {M{\'e}sz{\'a}ros} P.,  2020, \mn@doi [\apj]
  {10.3847/1538-4357/abbe00}, \href
  {https://ui.adsabs.harvard.edu/abs/2020ApJ...904..188K} {904, 188}

\bibitem[\protect\citeauthoryear{{Kulsrud} \& {Pearce}}{{Kulsrud} \&
  {Pearce}}{1969}]{1969ApJ...156..445K}
{Kulsrud} R.,  {Pearce} W.~P.,  1969, \mn@doi [\apj] {10.1086/149981}, \href
  {https://ui.adsabs.harvard.edu/abs/1969ApJ...156..445K} {156, 445}

\bibitem[\protect\citeauthoryear{{LHAASO Collaboration}}{{LHAASO
  Collaboration}}{2024}]{2024arXiv241008988L}
{LHAASO Collaboration} 2024, \mn@doi [arXiv e-prints]
  {10.48550/arXiv.2410.08988}, \href
  {https://ui.adsabs.harvard.edu/abs/2024arXiv241008988L} {p. arXiv:2410.08988}

\bibitem[\protect\citeauthoryear{{Lerche}}{{Lerche}}{1967}]{1967ApJ...147..689L}
{Lerche} I.,  1967, \mn@doi [\apj] {10.1086/149045}, \href
  {https://ui.adsabs.harvard.edu/abs/1967ApJ...147..689L} {147, 689}

\bibitem[\protect\citeauthoryear{{Li} \& {Chen}}{{Li} \&
  {Chen}}{2010}]{2010MNRAS.409L..35L}
{Li} H.,  {Chen} Y.,  2010, \mn@doi [\mnras]
  {10.1111/j.1745-3933.2010.00944.x}, \href
  {https://ui.adsabs.harvard.edu/abs/2010MNRAS.409L..35L} {409, L35}

\bibitem[\protect\citeauthoryear{{Malkov}, {Diamond}, {Sagdeev}, {Aharonian}
  \& {Moskalenko}}{{Malkov} et~al.}{2013}]{2013ApJ...768...73M}
{Malkov} M.~A.,  {Diamond} P.~H.,  {Sagdeev} R.~Z.,  {Aharonian} F.~A.,
  {Moskalenko} I.~V.,  2013, \mn@doi [\apj] {10.1088/0004-637X/768/1/73}, \href
  {https://ui.adsabs.harvard.edu/abs/2013ApJ...768...73M} {768, 73}

\bibitem[\protect\citeauthoryear{{Mattia}, {Del Zanna}, {Bugli}, {Pavan},
  {Ciolfi}, {Bodo}  \& {Mignone}}{{Mattia} et~al.}{2023}]{2023A&A...679A..49M}
{Mattia} G.,  {Del Zanna} L.,  {Bugli} M.,  {Pavan} A.,  {Ciolfi} R.,  {Bodo}
  G.,   {Mignone} A.,  2023, \mn@doi [\aap] {10.1051/0004-6361/202347126},
  \href {https://ui.adsabs.harvard.edu/abs/2023A&A...679A..49M} {679, A49}

\bibitem[\protect\citeauthoryear{{Morikawa}, {Ohira}  \& {Ohmura}}{{Morikawa}
  et~al.}{2024}]{2024ApJ...969L...1M}
{Morikawa} K.,  {Ohira} Y.,   {Ohmura} T.,  2024, \mn@doi [\apjl]
  {10.3847/2041-8213/ad50a2}, \href
  {https://ui.adsabs.harvard.edu/abs/2024ApJ...969L...1M} {969, L1}

\bibitem[\protect\citeauthoryear{{O'Sullivan}, {Reville}  \&
  {Taylor}}{{O'Sullivan} et~al.}{2009}]{2009MNRAS.400..248O}
{O'Sullivan} S.,  {Reville} B.,   {Taylor} A.~M.,  2009, \mn@doi [\mnras]
  {10.1111/j.1365-2966.2009.15442.x}, \href
  {https://ui.adsabs.harvard.edu/abs/2009MNRAS.400..248O} {400, 248}

\bibitem[\protect\citeauthoryear{{Ohira}}{{Ohira}}{2013}]{2013ApJ...767L..16O}
{Ohira} Y.,  2013, \mn@doi [\apjl] {10.1088/2041-8205/767/1/L16}, \href
  {http://adsabs.harvard.edu/abs/2013ApJ...767L..16O} {767, L16}

\bibitem[\protect\citeauthoryear{{Ohira} \& {Ioka}}{{Ohira} \&
  {Ioka}}{2011}]{2011ApJ...729L..13O}
{Ohira} Y.,  {Ioka} K.,  2011, \mn@doi [\apjl] {10.1088/2041-8205/729/1/L13},
  \href {http://adsabs.harvard.edu/abs/2011ApJ...729L..13O} {729, L13}

\bibitem[\protect\citeauthoryear{{Ohira}, {Murase}  \& {Yamazaki}}{{Ohira}
  et~al.}{2010}]{2010A&A...513A..17O}
{Ohira} Y.,  {Murase} K.,   {Yamazaki} R.,  2010, \mn@doi [\aap]
  {10.1051/0004-6361/200913495}, \href
  {http://adsabs.harvard.edu/abs/2010A%26A...513A..17O} {513, A17}

\bibitem[\protect\citeauthoryear{{Ohira}, {Murase}  \& {Yamazaki}}{{Ohira}
  et~al.}{2011}]{2011MNRAS.410.1577O}
{Ohira} Y.,  {Murase} K.,   {Yamazaki} R.,  2011, \mn@doi [\mnras]
  {10.1111/j.1365-2966.2010.17539.x}, \href
  {http://adsabs.harvard.edu/abs/2011MNRAS.410.1577O} {410, 1577}

\bibitem[\protect\citeauthoryear{{Ohira}, {Yamazaki}, {Kawanaka}  \&
  {Ioka}}{{Ohira} et~al.}{2012}]{2012MNRAS.427...91O}
{Ohira} Y.,  {Yamazaki} R.,  {Kawanaka} N.,   {Ioka} K.,  2012, \mn@doi
  [\mnras] {10.1111/j.1365-2966.2012.21908.x}, \href
  {https://ui.adsabs.harvard.edu/abs/2012MNRAS.427...91O} {427, 91}

\bibitem[\protect\citeauthoryear{{Ohira}, {Kisaka}  \& {Yamazaki}}{{Ohira}
  et~al.}{2018}]{2018MNRAS.478..926O}
{Ohira} Y.,  {Kisaka} S.,   {Yamazaki} R.,  2018, \mn@doi [\mnras]
  {10.1093/mnras/sty1159}, \href
  {https://ui.adsabs.harvard.edu/abs/2018MNRAS.478..926O} {478, 926}

\bibitem[\protect\citeauthoryear{{Ohmura}, {Ono}, {Sakemi}, {Tashima}, {Omae}
  \& {Machida}}{{Ohmura} et~al.}{2021}]{2021ApJ...910..149O}
{Ohmura} T.,  {Ono} K.,  {Sakemi} H.,  {Tashima} Y.,  {Omae} R.,   {Machida}
  M.,  2021, \mn@doi [\apj] {10.3847/1538-4357/abe5a1}, \href
  {https://ui.adsabs.harvard.edu/abs/2021ApJ...910..149O} {910, 149}

\bibitem[\protect\citeauthoryear{{Ostrowski}}{{Ostrowski}}{1998}]{1998A&A...335..134O}
{Ostrowski} M.,  1998, \mn@doi [\aap] {10.48550/arXiv.astro-ph/9803299}, \href
  {https://ui.adsabs.harvard.edu/abs/1998A&A...335..134O} {335, 134}

\bibitem[\protect\citeauthoryear{{Pacini} \& {Salvati}}{{Pacini} \&
  {Salvati}}{1973}]{1973ApJ...186..249P}
{Pacini} F.,  {Salvati} M.,  1973, \mn@doi [\apj] {10.1086/152495}, \href
  {https://ui.adsabs.harvard.edu/abs/1973ApJ...186..249P} {186, 249}

\bibitem[\protect\citeauthoryear{{Porth}, {Komissarov}  \& {Keppens}}{{Porth}
  et~al.}{2014}]{2014MNRAS.438..278P}
{Porth} O.,  {Komissarov} S.~S.,   {Keppens} R.,  2014, \mn@doi [\mnras]
  {10.1093/mnras/stt2176}, \href
  {https://ui.adsabs.harvard.edu/abs/2014MNRAS.438..278P} {438, 278}

\bibitem[\protect\citeauthoryear{{Rieger} \& {Duffy}}{{Rieger} \&
  {Duffy}}{2004}]{2004ApJ...617..155R}
{Rieger} F.~M.,  {Duffy} P.,  2004, \mn@doi [\apj] {10.1086/425167}, \href
  {https://ui.adsabs.harvard.edu/abs/2004ApJ...617..155R} {617, 155}

\bibitem[\protect\citeauthoryear{{Romero} \& {Vila}}{{Romero} \&
  {Vila}}{2008}]{2008A&A...485..623R}
{Romero} G.~E.,  {Vila} G.~S.,  2008, \mn@doi [\aap]
  {10.1051/0004-6361:200809563}, \href
  {https://ui.adsabs.harvard.edu/abs/2008A&A...485..623R} {485, 623}

\bibitem[\protect\citeauthoryear{{Sakemi}, {Omae}, {Ohmura}  \&
  {Machida}}{{Sakemi} et~al.}{2021}]{2021PASJ...73..530S}
{Sakemi} H.,  {Omae} R.,  {Ohmura} T.,   {Machida} M.,  2021, \mn@doi [\pasj]
  {10.1093/pasj/psab018}, \href
  {https://ui.adsabs.harvard.edu/abs/2021PASJ...73..530S} {73, 530}

\bibitem[\protect\citeauthoryear{{Thoudam}, {Rachen}, {van Vliet},
  {Achterberg}, {Buitink}, {Falcke}  \& {H{\"o}randel}}{{Thoudam}
  et~al.}{2016}]{2016A&A...595A..33T}
{Thoudam} S.,  {Rachen} J.~P.,  {van Vliet} A.,  {Achterberg} A.,  {Buitink}
  S.,  {Falcke} H.,   {H{\"o}randel} J.~R.,  2016, \mn@doi [\aap]
  {10.1051/0004-6361/201628894}, \href
  {https://ui.adsabs.harvard.edu/abs/2016A&A...595A..33T} {595, A33}

\bibitem[\protect\citeauthoryear{{Torres}, {Rodriguez Marrero}  \& {de Cea Del
  Pozo}}{{Torres} et~al.}{2008}]{2008MNRAS.387L..59T}
{Torres} D.~F.,  {Rodriguez Marrero} A.~Y.,   {de Cea Del Pozo} E.,  2008,
  \mn@doi [\mnras] {10.1111/j.1745-3933.2008.00485.x}, \href
  {https://ui.adsabs.harvard.edu/abs/2008MNRAS.387L..59T} {387, L59}

\bibitem[\protect\citeauthoryear{{Weaver}, {McCray}, {Castor}, {Shapiro}  \&
  {Moore}}{{Weaver} et~al.}{1977}]{1977ApJ...218..377W}
{Weaver} R.,  {McCray} R.,  {Castor} J.,  {Shapiro} P.,   {Moore} R.,  1977,
  \mn@doi [\apj] {10.1086/155692}, \href
  {https://ui.adsabs.harvard.edu/abs/1977ApJ...218..377W} {218, 377}

\bibitem[\protect\citeauthoryear{{Zabalza}}{{Zabalza}}{2015}]{2015ICRC...34..922Z}
{Zabalza} V.,  2015, in 34th International Cosmic Ray Conference (ICRC2015).
  p.~922 (\mn@eprint {arXiv} {1509.03319}), \mn@doi{10.22323/1.236.0922}

\makeatother
\end{thebibliography}








\bsp	
\label{lastpage}
\end{document}